\documentclass[12pt]{article}

\usepackage{epsfig}

\def\be{\begin{equation}}
\def\ee{\end{equation}}
\def\beq{\begin{eqnarray}}
\def\eeq{\end{eqnarray}}

\def\gsim{\:\raisebox{-0.5ex}{$\stackrel{\textstyle>}{\sim}$}\:} 

\begin{document}
\begin{flushright}
TIFR/TH/04-25
\end{flushright}
\bigskip

\begin{center}
{\Large{\bf Solar Neutrino Oscillation -- An Overview$^\dagger$}} \\[3cm]
{\large D.P. Roy} \\[1cm]
Tata Institute of Fundamental Research, \\ Homi Bhabha Road, 
Mumbai 400 005, India
\end{center}
\bigskip\bigskip

After a brief summary of the neutrino oscillation formalism and the
solar neutrino sources and experiments I discuss the matter effect on
solar neutrino oscillation.  Then I discuss how the resulting
alternative solutions are experimentally resolved in favour of the LMA
solution, with particular exphasis on the SK, SNO and KL data.

\vfill

\noindent $^\dagger$Plenary Talk at the Xth International Symposium on
Particle, Strings and Cosmology (PASCOS), Boston, 16-22 August 2004.

\newpage

The last four years have been widely described as the golden years of
solar neutrino physics, thanks to three pioneering experiments --
SuperKamiokande (SK), Solar Neutrino Observatory (SNO) and KamLAND
(KL).  They have provided for the first time a unique solution to the
solar neutrino anomaly in terms of neutrino oscillation with
unambiguous mass and mixing parameters.  I shall give an overview of
the subject with particular emphasis on the role of these
experiments.  After a brief summary of the neutrino oscillation
formalism and the solar neutrino sources and experiments I shall
discuss the matter effect on solar neutrino oscillation.  We shall see
how it leads to four alternative solutions to the solar neutrino
anomaly and then their experimental resolution in favour of the 
so called Large Mixing Angle (LMA) solution over the last four years.
\bigskip

\noindent {\bf Neutrino Mixing and Oscillation:} It was already noted by
Pontecorvo back in the sixties that if the neutrinos have non-zero
mass then there will in general be mixing between the flavour and the mass
eigenstates, which will lead to neutrino oscillation [1].  We assume
for simplicity two neutrino flavours, in which case mixing can be
discribed by one angle $\theta$, i.e.
\be
\left(\matrix{\nu_e \cr \nu_\mu}\right) = \left(\matrix{\cos\theta &
\sin\theta \cr -\sin\theta & \cos\theta}\right) \left(\matrix{\nu_1
\cr \nu_2}\right).
\label{one}
\ee
In fact it provides a very good approximation to the three neutrino
mixing scenario for solar neutrino oscillation, with the second
neutrino representing a mixture of the $\nu_\mu$ and $\nu_\tau$
flavours.  Now each mass eigenstate propagates with its own phase
\be
e^{-i(Et - p\ell)} \simeq e^{-{im^2\ell \over 2E}}, 
\label{two}
\ee
where we have made the relativistic approximation, $E = p + m^2/2p$,
since neutrino masses are much smaller than their kinetic energy.
Thus a $\nu_e$ produced at the origin will propagate as
\be
\nu_e \rightarrow \nu_1 \cos\theta e^{-{im^2_1 \over 2E} \ell} + \nu_2
\sin\theta e^{-{im^2_2 \over 2E} \ell}.
\label{three}
\ee
Decomposing the $\nu_{1,2}$ back into $\nu_{e,\mu}$ after a distance
$\ell$ one sees that the $\nu_\mu$ terms do not cancel, which implies
neutrino oscillation.  In fact the coefficient of the $\nu_\mu$ term
represents the probability of $\nu_e \rightarrow \nu_\mu$ oscillation,
i.e. 
\beq
P_{\nu_e \rightarrow \nu_\mu} (\ell) &=& \left|\cos\theta \sin\theta
\left(e^{-{im^2_2 \over 2E} \ell} - e^{-{im^2_1 \over 2E}\ell} \right)
\right|^2 \nonumber \\[2mm]
&=& \sin^2 2\theta \sin^2 \left({\Delta m^2 \over 4E} \ell \right), 
\label{four}
\eeq
where the first factor represents the amplitude and the second factor
the phase of oscillation, with $\Delta m^2 = m^2_1 - m^2_2$.
Converting to convenient units for $\Delta m^2 (eV^2), \ \ell (m)$ and
$E \ ({\rm MeV})$ one gets
\be
P_{\nu_e \rightarrow \nu_\mu} (\ell) = \sin^2 2\theta \sin^2 (1.3
\Delta m^2 \ell/E),
\label{five}
\ee
which corresponds to an oscillation wavelength
\be
\lambda = {\pi \over 1.3} \cdot {E \over \Delta m^2} \simeq 2.4
E/\Delta m^2.
\label{six}
\ee

Thus for large mixing angle $(\sin^2 2\theta \sim 1)$ one expects the
following pattern of oscillation probability from eqs. (\ref{five})
and (\ref{six}), where the factor of ${1\over2}$ in the last case
comes from averaging over the phase factor.
\be
\matrix{
\ell \ \ \ & \ \ \ \ll \lambda \ \ \ & \ \ \ = \lambda/2 \ \ \ & \ \ \
\gg \lambda \cr \cr  
P_{\nu_e \rightarrow \nu_\mu} \ \ \ & \ \ \ 0 \ \ \ & \ \ \ \sin^2
2\theta \sim 1 \ \ \ & \ \ \ {1\over2} \sin^2 2\theta \sim {1\over2}}
\label{seven} 
\ee
Note that the corresponding survival probability is given by the
remainder, i.e.
\be
P_{ee} \equiv P_{\nu_e \rightarrow \nu_e} = 1 - P_{\nu_e \rightarrow
\nu_\mu}. 
\label{eight}
\ee
Now the typical energy of $\nu_e$ coming from a nuclear reactor or the
sun is $\sim 1$ MeV.  The typical distance between the source and the
dector is a few hundred Km $(\ell \sim 10^5 \ {\rm m})$ for the long
base line KamLAND reactor experiment, while $\ell \sim 10^{11} \ {\rm
m}$ for solar neutrino experiments.  Thus one sees from
eqs. (\ref{six}) and (\ref{seven}) that the KamLAND and the solar
neutrino experiments can probe neutrino mass down to $\Delta m^2 \sim
10^{-5} \ {\rm eV}^2$ and $10^{-11} \ {\rm eV}^2$ respectively, which
is far beyond the reach of any other method of mass measurement.
\bigskip

\noindent {\bf Solar Neutrino Sources and Experiments:} The main sources of
solar neutrinos are the $pp$ chains of nuclear reaction, taking place
at the solar core, which convert protons into $^4He$ ($\alpha$
particle).  They are
\begin{enumerate}
\item[{(I)}] $pp \rightarrow {^2H} + e^+ + \nu_e$, ${^2H} + p
\rightarrow \ {^3He} + \gamma$, ${^3He} + {^3He} \rightarrow {^4He} + 2p$; or
\item[{(II)}] ${^3He} + {^4He} \rightarrow {^7Be} + \gamma$, ${^7Be} + e^-
\rightarrow {^7Li} + \nu_e$, ${^7Li} + p \rightarrow 2 \ {^4He}$; or
\item[{(III)}] ${^7Be} + p \rightarrow {^8B} + \gamma$, ${^8B} \rightarrow
{^8Be^\star} + e^+ + \nu_e$, ${^8Be^\star} \rightarrow 2 \ {^4He}$.
\end{enumerate}
While most of this conversion takes place by the shortest path (I) a
small fraction (15\%) follows a detour via ${^7Be}$ (II); and a tiny
fraction (0.1\%) of the latter follows a still longer detour via
${^8B}$ (III).  The resulting neutrinos are (I) the low energy $pp$
neutrino, (II) the intermediate energy $Be$ neutrino and (III) the
relatively high energy $B$ neutrino, with decreasing order of flux.
The standard solar model (SSM) prediction for these fluxes is shown in
Fig. 1 from BP 2000 [2].  It also shows the neutrino energy ranges
covered by the different solar neutrino experiments.

The Gallium [3] and the Chlorine [4] experiments are based on the
charged current reactions
\be
\nu_e + {^{71}Ga} \rightarrow e^- + {^{71}Ge} \ \ {\rm and} \ \ \nu_e
+ {^{37}Cl} \rightarrow e^- + {^{37}Ar}.
\label{nine}
\ee
The produced ${^{71}Ge}$ and ${^{37}Ar}$ are periodically extracted by
radiochemical method, from which the incident neutrino fluxes are
estimated.  The value of measured flux $R$ relative to the SSM
prediction gives the $\nu_e$ survival probability $P_{ee}$.  The SK is
a real-time water Cerenkov experiment [5], based on the elastic
scattering of neutrino on electron.  The elastic scattering is
dominated by the charged current reaction $\nu_e + e^- \ {\buildrel CC
\over \longrightarrow} \ \nu_e + e^-$, but it also has a limited
sensitivity to the neutral current reaction $\nu_{e,\mu} + e^- \
{\buildrel NC \over \longrightarrow} \ \nu_{e,\mu} + e^-$. Thus
\be
R_{e\ell} = P_{ee} + {\sigma^{NC} \over \sigma^{NC+CC}} (1 - P_{ee})
\simeq P_{ee} + {1\over6} (1 - P_{ee}).
\label{ten}
\ee
This experiment can also measure the energy and direction of the
incident neutrino from those of the outgoing electron.  The SNO is a
Cerenkov experiment with a heavy water target, which can detect both
the charged and neutral current events [6-9]
\be
\nu_e + d \ {\buildrel CC \over \longrightarrow} \ p + p + e^-,
\label{eleven}
\ee
\be
\nu_{e,\mu} + d \ {\buildrel NC \over \longrightarrow} \ \nu_{e,\mu} +
p + n.
\label{twelve}
\ee
In the 1st phase of the experiment the $NC$ events were detected via
neutron capture on deuteron, $n + d \rightarrow t + \gamma$ [7].  In
the 2nd phase salt was added to the heavy water target to enhance the
$NC$ detection efficiency via neutron capture on chlorine, $n +
{^{35}Cl} \rightarrow {^{36}Cl} + \gamma$ [8].  In the 3rd phase of
this experiment going on now $^3He$ gas filled counters are inserted
into the heavy water target to detect $NC$ events via $n + {^3He}
\rightarrow t + p$ [9].

Table 1 shows the energy thresholds of the above four experiments
along with the compositions of the corresponding solar neutrino
spectra.  It also shows the corresponding survival probability
$P_{ee}$ measured by the rates of the charged current reactions
relative to the SSM prediction [2].  For the $SK$ experiment the
survival probability calculated from $R_{e\ell}$ via eq. (\ref{ten})
is shown in parentheses.  It shows that the survival probability is
slightly above 1/2 for low energy $\nu_e$, falling to 1/3 at high
energy.  To understand its magnitude and the energy dependence we have
to consider the effect of solar matter on neutrino oscillation.
\bigskip

\noindent {\bf Table 1.} The $\nu_e$ survival probability $P_{ee}$
measured by the $CC$ event rates of various solar neutrino experiments
relative to the SSM prediction.  For SK the $P_{ee}$ obtained after NC
correction is shown in bracket.  The energy threshold and composition
of neutrino beam are also shown for each experiment.
\[
{\small
\begin{tabular}{|c|c|c|c|c|}
\hline
Experiment & Gallium & Chlorine & SK & SNO-I \\
\hline
R & $0.55 \pm 0.03$ & $0.33 \pm 0.03$ & $0.465 \pm 0.015$ & $0.35 \pm
0.03$ \\
  &                 &                 & $(0.36 \pm 0.015)$ & \\
\hline
$E_{th}$ (MeV)      & $0.2$           & $0.8$ & $5$ & $5$ \\ 
\hline
Composition & $pp$ (55\%),  & $B$ (75\%), & $B$ (100\%) & $B$ (100\%) \\
            & $Be$ (25\%), $B$ (10\%) & $Be$ (15\%)  && \\
\hline
\end{tabular}
}
\]
\bigskip

\noindent {\bf Matter Enhancement (Resonant Conversion):} Propagation
through solar matter gives an induced mass to $\nu_e$, which can have
profound effect on neutrino oscillation.  This is known as MSW effect
after its authors [10].  It arises from the charged current
interaction of $\nu_e$ with solar electrons, while the neutral current
interaction has no net effect since it is flavour independent. Adding
this interaction energy density to the free particle wave equation
gives 
\be
-i {d \over dt} \left(\matrix{\nu_e \cr \nu_\mu}\right) = \left(p +
{M^2 + 2E H_{int} \over 2E}\right) \left(\matrix{\nu_e \cr
\nu_\mu}\right), 
\label{thirteen}
\ee    
where $H_{int} = \sqrt{2} G N_e (0)$ for $\nu_e (\nu_\mu)$, with $G$
and $N_e$ denoting Fermi coupling and solar electron density.  The
quantity on the rhs numerator can be regarded as an effective mass or
energy, i.e.
\beq
M^{\prime 2} &=& \left(\matrix{c & s \cr -s & c}\right)
\left(\matrix{m^2_1 & 0 \cr 0 & m^2_2}\right) \left(\matrix{c & -s \cr
s & c}\right) + \left(\matrix{2\sqrt{2} E G N_e & 0 \cr 0 & 0}\right)
\nonumber \\[2mm] 
&=& \left(\matrix{c^2 m^2_1 + s^2 m^2_2 + 2\sqrt{2} E G N_e & -sc
\Delta m^2 \cr -sc\Delta m^2 & c^2 m^2_2 + s^2 m^2_1}\right),
\label{fourteen}
\eeq
where $s,c$ denote $\sin\theta,\cos\theta$.

To get a simple picture let us assume for the moment that $\sin\theta
\ll 1$, so that the nondiagonal elements are small.  Then we can
identify the two diagonal elements with the eigenvalues and the
corresponding eigenstates $\nu_{1,2}$ with the flavour eigenstates
$\nu_{e,\mu}$ respectively.  Fig. 2 shows the two eigenvalues
$\lambda_{1,2}$ against the solar electron density.  At the solar
surface $(N_e = 0)$ the 1st eigenvalue $(m^2_1)$ is smaller than the
2nd $(m^2_2)$.  But $\lambda_1$ increases steadily with $N_e$ and
becomes much larger than $\lambda_2$ at the solar core.  The
cross-over occurs at a critical density
\be
N^c_e \equiv {\Delta m^2 \over 2\sqrt{2} GE} \cos 2\theta,
\label{fifteen}
\ee
corresponding to $M^{\prime 2}_{11} = M^{\prime 2}_{22}$.  Note
however that the two eigenvalues actually never cross.  There is a
minimum gap between them given by the nondiagonal element, i.e.
\be
\Gamma = \Delta m^2 \sin^2 2\theta.
\label{sixteen}
\ee
This means that a $\nu_e$ produced at the solar core will come out as
$\nu_2$, provided the transition probability between the two energy
levels remains small.

It is easy to show that this remarkable result does not depend on the
$\sin\theta \ll 1$ assumption.  Consider the effective mixing angle
$\theta_M$ in matter, which diagonalises the above matrix, i.e.
\be
\tan 2\theta_M = {2M^{\prime 2}_{12} \over |M^{\prime 2}_{22} -
M^{\prime 2}_{11}|} = {\sin 2\theta \over |\cos2\theta - 2\sqrt{2} G E
N_e/\Delta m^2|}.
\label{seventeen}
\ee
The electron density at the solar core is $N^0_e \gg N^c_e$, so that
the 2nd term in the denominator of eq. (\ref{seventeen}) is much
larger than the 1st.  This means $\theta_M \ll 1$ at the solar core
for any vacuum mixing angle $\theta$; so that the $\nu_e$ produced
there is dominated by the $\nu_1$ component.  At the critical density
the denominator of eq. (\ref{seventeen}) vanishes, which corresponds
to maximal mixing between the two components, again for any value of
$\theta$.  This is why it is called matter enhanced (or resonant)
conversion.  It comes out from the sun as $\nu_2$ with
\be
P_{ee} = \sin^2 \theta,
\label{eighteen}
\ee
provided the transition probability between the two levels remains
small through out the propagation and in particular in the critical
density region.  This transition probability is given by the
Landau-Zenner formula, i.e.
\be
T = e^{-{\pi \over 2\gamma}}, \ \gamma = {\lambda (d\lambda_1/d\ell)_c
\over \Gamma} \propto \lambda {d N_e/d\ell \over N_e}\Big|_c,
\label{ninteen}
\ee
where $\lambda$ represents the oscillation wavelength in matter (which
should not be confused with the eigenvalues $\lambda_{1,2}$).  If the
solar electron density varies so slowly that the resulting variation
in the 1st eigenvalue over an oscillation wavelength is small compared
to the gap between the two, then $\gamma \ll 1$ and the transition
rate is exponentially suppressed.  This is called the adiabatic
condition.  Thus the two conditions for the solar $\nu_e$ to emerge as
$\nu_2$ can be written as
\be
{\Delta m^2 \cos 2\theta \over 2\sqrt{2} G N^0_e} < E < {\Delta m^2
\sin^2 2\theta \over 2\cos 2\theta (d N_e/d\ell N_e)_c}, 
\label{twenty}
\ee
where the 1st inequality ensures $N^0_e > N^c_e$ andthe 2nd one is the
adiabatic condition.

Fig. 3 shows the triangular region in the $\Delta m^2 - \sin^2
2\theta$ plot satisfying the above two conditions for a typical
neutrino energy of $E = 1$ MeV.  The horizontal side follows from the
1st inequality, which gives a practically constant upper limit of
$\Delta m^2$ since $\cos 2\theta \simeq 1$.  The 2nd inequality
(adiabatic condition) gives a lower limit on $\sin^2 2\theta$.
Moreover since this condition implies a lower limit on the product
$\Delta m^2 \sin^2 2\theta$, it corresponds to the diagonal line on
the log-log plot.  The vertical side is simply given by the physical
boundary, corresponding to maximal mixing.  This is the socalled MSW
triangle.  The indicated survival probabilities out side the triangle
follow from the vacuum oscillation formulae (\ref{seven}) and
(\ref{eight}), while that inside is given by eq. (\ref{eighteen}).
Thus $P_{ee} < 1/2$ inside the MSW triangle and $> 1/2$ outside it,
except for the oscillation maximum at the bottom $(\ell \simeq
\lambda/2)$ where $P_{ee}$ goes down to $\cos^2 2\theta$.  Finally the
earth matter effect gives a small but positive $\nu_e$ regeneration
probability, which implies a day-night a symmetry -- i.e. the sun
shines a little brighter at night in the $\nu_e$ beam.  After
day-night averaging one expects a $\nu_e$ regeneration probability
\be
\delta P_{reg} = {\eta_E \sin^2 2\theta \over 4(1 - 2\eta_E \cos
2\theta + \eta^2_E)}, \ \ \eta_E = {0.66 \over \rho Y_e} \left({\Delta
m^2/E \over 10^{-13} \ {\rm eV}}\right),
\label{twentyone}
\ee
where $\rho$ is matter density in the earth in $gm/cc$ and $Y_e$ the
average number of electrons per nucleon.  For favourable values of
$\Delta m^2$ and $\theta \ \delta P_{reg}$ can go upto 0.15 as indicated
in the Figure.  It is important to note that the positions of the MSW
triangle, the earth regeneration region and the vacuum oscillation
maximum depend only on the ratio $\Delta m^2/E$, as one can see from
the relevant formulae.  Thus their positions on the right hand scale
hold at all energies. 
\bigskip

\noindent {\bf Four Alternative Solutions:} Fig. 3 marks four regions
in the mass and mixing parameter space, which can explain the
magnitude and energy dependence of the survival probability $P_{ee}$
shown in Table 1.  They correspond to the socalled Large Mixing Angle
(LMA), Small Mixing Angle (SMA), Low Mass (LOW) and Vacuum Oscillation
(VAC) solutions.  For the LMA and SMA solutions $(\Delta m^2 \sim
10^{-5} \ {\rm eV}^2)$ the low energy $Ga$ experiment $(E \ll 1 \ {\rm
MeV})$ falls above the MSW triangle in $\Delta m^2/E$, while the SK
and SNO experiments $(E \gg 1 \ {\rm MeV})$ fall inside it.  Therefore
the solar matter effect can explain the observed decrease of the
survial probability with increasing energy.  For the LOW solution the
low energy $Ga$ experiment is pushed up to the region indicated by the
dashed line, where it gets an additional contribution to the survival
probability from the earth regeneration effect.  Finally the VAC
solution explains the energy dependence of the survival probability
via the energy dependence of the oscillation phase in
eq. (\ref{five}).  Fig. 4 shows the predicted survival probabilities
of the four solutions as functions of neutrino energy.  The LMA and
LOW solutions predict mild and monotonic energy dependence, while the
SMA and VAC solutions predict very strong and nonmonotonic energy
dependence. 
\bigskip

\noindent {\bf Experimental Resolution in Favour of LMA:} The survival
rates in Table 1 show a slight preference for a nonmonotonic energy
dependence, since the intermediate energy Chlorine experiment shows a
little lower rate than SK and SNO.  Therefore the SMA and VAC were the
favoured solutions in the early days.  However the situation changed
drastically with the measurement of the energy spectrum by SK [5],
shown in Fig. 5.  It shows very little energy depence in clear
disagreement with the predictions of the SMA and VAC solutions in
Fig. 4.  In particular the SMA solution in Fig. 4 can not reconcile a
low survival rate of $P_{ee} \simeq 1/3$ with an energy independent
spectrum, in conflict with the SK data.  The first charged current
data from SNO [6] agreed with the low survival rate as well as the
energy independent spectrum of the SK data.  Thus the global analyses
of the solar neutrino data at this stage ruled out the SMA and VAC
solutions in favour of LMA and LOW [11].

Then came the first neutral current data from SNO [7].  Being flavour
independent the neutral current reaction is unaffected by neutrino
oscillation.  Hence it can be used to measure the Boron neutrino flux,
for which the SSM has a large uncertainty (see Fig. 1).  The SNO
neutral current measurement of this flux was in agreement with the SSM
prediction and significantly more precise than the latter.  Fig. 6
shows the results of global fit with and without the SNO (NC) data
[12].  The left and the middle panels show two different methods of
using the NC data and give very similar results.  They strongly favour
the LMA solution while barely allowing LOW at $3\sigma$ level.  The
reason is that the earth regeneration contribution is too small to
account for the large survival rate of the $Ga$ compared to those
of the SK and SNO (CC) experiments.  Without the SNO (NC) data it was
possible to effectively push up the survival rate of the latter
experiments from 0.35 to 0.45 by exploiting the large uncertainty in
the Boron neutrino flux, and hence accommodate the LOW solution as
shown in the right pannel (see also Fig. 4).  However this was no
longer possible after the SNO (NC) data [12,13] due to the improved
precission of the Boron neutrino flux.
\bigskip

\noindent {\bf Confirmation and Sharpening of the LMA Solution:}
Independent confirmation of the LMA solution came from the reactor
antineutrino data of the KamLAND experiment [14], assuming CPT
invariance.  It is a 1 kiloton liquid scintillator experiment
detecting $\bar{\nu_e}$ from the Japanese nuclear reactors via
\be
\bar{\nu_e} + p \rightarrow e^+ + n.
\label{twentytwo}
\ee
It also measures the incident $\bar{\nu_e}$ energy $E$ via the visible
scintillation energy produced by the positron and its annihilation
with a target electron, i.e.
\be
E_{vis} = E + m_e + m_p - m_n = 
E - 0.8 \ {\rm MeV}.
\label{twentythree}
\ee
The mean base line distance of the detector from the reactors is
$\langle \ell \rangle \sim 180$ km, which means it is sensitive to the
$\Delta m^2 \gsim 10^{-5} \ {\rm eV}^2$ region as mentioned earlier.
Thus the experiment was designed to probe the LMA region.  It was
shown in [15] that if the survival rate seen at KL is $< 0.9$, then it
would rule out the LOW solution at $3\sigma$ level.  The first KL
result from 162ty data showed a survival rate $R = 0.611 \pm .085 \pm
.041$ [14].  This was in perfect agreement with the LMA prediction and
ruled out LOW at $5\sigma$ level.  Moreover the observed spectral
distortion of the KL data, taken together with global solar neutrino
data, confined the LMA solution to two subregions around $\Delta m^2 =
7$ and $14 \times 10^{-5} \ {\rm eV}^2$, as shown in Fig. 7 [16].
They correspond to the 1st and 2nd oscillation minima ($\langle \ell
\rangle = \lambda$ and $2\lambda$) for $E_{vis} \simeq 4.5 \ {\rm
MeV}$, where the observed spectrum touches the no-oscillation
prediction (see Fig. 9).  The best fit point lies in the lower region
called LMA-I, while LMA-II is allowed only at 99\% CL [16,17].

This was followed by the data from the second (salt) phase of SNO [8],
with better $NC$ detection efficiency.  Combining the data from the
two phases in a global analysis helped to constrain the mass and
mixing parameters further [18,19].  Fig. 8 shows the results of global
fits with phase-1, phase-2 and the combined SNO data [18].  The
combined fit is seen to allow the LMA-II region only at $3\sigma$ and
disallow maximal mixing at $5\sigma$.  The most important issue at
this point was a definitive resolution of the LMA-I and II ambiguity.
It was shown through a simulation study in [18] that if the KL
spectrum from 1 kty data continues to favour the LMA-I region, then
combining this with the global solar neutrino data will rule out
LMA-II at $> 3\sigma$ level.

Recently the KL experiment has published their 766 ty data, whose best
fit point is indeed in the LMA-I region [20].  Combining this with the
global solar neutrino data rules out LMA-II at $> 3\sigma$ while
sharpening the LMA-I region further [21,22].  Fig. 9 compares the KL
spectrum with the no-oscillation and oscillation best fit predictions.
Fig. 10 shows the result of combined fit to the KL and global solar
neutrino data [21].  The 90\% CL contour of the global solar fit is
shown for comparison.  In comparing the two 90\% CL contours we see
that the mass parameter is mainly determined by the KL data, while the
mixing angle is determined mainly by the solar neutrino data.  The
best fit values along with $1\sigma$ errors are
\be
\Delta m^2 = (8.0 \pm .6) \times 10^{-5} \ {\rm eV}^2, \ \sin^2 \theta
= 0.28 \pm .03.
\label{twentyfour}
\ee
The best fit value of $\Delta m^2$ corresponds to the 1st oscillation
minimum $(\lambda = \langle \ell \rangle = 180 km)$ occuring at
$E_{vis} \simeq 5$ MeV, in agreement with Fig. 9.  It is evident from
eq. (\ref{twentyfour}) and Fig. 10 that the solar neutrino oscillation
has finally entered the arena of precission physics.
\bigskip

\noindent {\bf Concluding Remarks:} Oscillation analysis of global
solar neutrino plus KL data has been extended to three neutrino
flavours [21].  The resulting mass and mixing parameters agree very
well with those of the two-flavour analysis discussed above.  Moreover
dropping one of the solar neutrino experiments from the global
analysis is also found to make little difference to the result.  Thus
the results are stable and robust.  As regards future prospects, the
precission of $\Delta m^2$ will improve further with accumulation of
more KL data.  One expects some improvement in the precission of the
mixing angle from the 3rd phase of SNO, but not from KL.  The main
reason for this is the occurence of an oscillation minimum in the
middle of the $\bar{\nu_e}$ spectrum $(E_{vis} \simeq 5 \ {\rm MeV})$.
This means that the coefficient of $\sin^2 2\theta$ in the oscillation
probability (eq. \ref{five}) is very small over this energy range.  An
interesting suggestion made in [23] is to reduce the base line length
by half $(\sim \lambda/2)$, which would mean an oscillation maximum in
the middle of the $\bar{\nu_e}$ spectrum instead.  A KL type
experiment at a reduced base line length of $\sim 70$ km has been
estimated to improve the precission of $\sin^2 \theta$ significantly.

It is my pleasure to thank the organisers of PASCOS'04 for their
invitation and kind hospitality.  Le me also take this opportunity to
thank my teammates Abhijit Bandyopadhyay, Sandhya Choubey and
Srubabati Goswami.

\newpage

\begin{center}
{\bf References}
\end{center}

\begin{enumerate}
\item[{1.}] B. Pontecorvo, Zh. Eksp. Teor. Fiz. 53 (1967) 1717.
\item[{2.}] J.N. Bahcall, M.H. Pinsonneault and S. Basu, Astrophys
J. 555 (2001) 990; http://www.sns.ias.edu/$\sim$jnb.
\item[{3.}] SAGE: J.N. Abdurashitov et al., JETP 95 (2002) 181; \\
GALLEX: W. Hampel et al., Phys. Lett. B447 (1999) 127; \\
GNO: C. Cattadori, Talk at Neutrino 2004, Paris (2004).
\item[{4.}] B.T. Cleveland et al., Astro Phys. J. 496 (1998) 505.
\item[{5.}] SK Collaboration: S. Fukuda et al., Phys. Lett. B539
(2002) 179.
\item[{6.}] SNO: Q.R. Ahmad et al., Phys. Rev. Lett. 87 (2001) 071301.
\item[{7.}] SNO: Q.R. Ahmad et al., Phys. Rev. Lett. 89 (2002) 011301
\& 011302.
\item[{8.}] SNO: S.N. Ahmed et. al., arXiv: nucl-ex/0309004.
\item[{9.}] SNO: H. Robertson, Talk at TAUP 2003, Seattle (2003).
\item[{10.}] L. Wolfenstein, Phys. Rev. D17 (1978) 2369; S.P. Mikheyev
and A.Y. Smirnov, Sov. J. Nucl. Phys. 42 (1985) 913 [Yad. Fiz. 42
(1985) 1441].
\item[{11.}] A. Bandyopadhyay, S. Choubey, S. Goswami and K. Kar,
Phys. Lett. B519 (2001) 83; G.L. Fogli, E. Lisi, D. Montanino and
A. Palazzo, Phys. Rev. D64 (2001) 093007; J.N. Bahcall,
M.C. Gonzalez-Garcia and C. Pena-Garay, JHEP 0108 (2001) 014;
P.I. Krastev and A.Y. Smirnov, Phys. Rev. D65 (2002) 073002.
\item[{12.}] A. Bandyopadhyay, S. Choubey, S. Goswami and D.P. Roy,
Phys. Lett. B540 (2002) 14.
\item[{13.}] V. Barger, D. Marfatia, K. Whisnant and B.P. Wood,
Phys. Lett. B537 (2002) 179; J.N. Bahcall, M.C. Gonzalez-Garcia and
C. Pena-Garay, JHEP 0207 (2002) 054; P.C. de Hollanda and
A.Y. Smirnov, Phys. Rev. D66 (2002) 113005; A. Strumia, C. Cattadori,
N. Ferrari and F. Vissani, Phys. Lett. B541 (2002) 327.
\item[{14.}] KamLAND: K. Eguchi et al., Phys. Rev. Lett. 90 (2003)
021802. 
\item[{15.}] A. Bandyopadhyay, S. Choubey, R. Gandhi, S. Goswami and
D.P. Roy, J. Phys. G29 (2003) 2465.
\item[{16.}] A. Bandyopadhyay, S. Choubey, R. Gandhi, S. Goswami and
D.P. Roy, Phys. Lett. B559 (2003) 121.
\item[{17.}] G.L. Fogli et al., Phys. Rev. D67 (2003) 073002;
M. Maltoni, T. Schwetz and J.W. Valle, Phys. Rev. D67 (2003) 093003.
J.N. Bahcall, M.C. Gonzalez-Garcia and C. Pena-Garay, JHEP 0302 (2003)
009; P.C. de Holanda and A.Y. Smirnov, JCAP 0302 (2003) 001.
\item[{18.}] A. Bandyopadhyay, S. Choubey, S. Goswami, S.T. Petcov and
D.P. Roy, Phys. Lett. B583 (2004) 134.
\item[{19.}] G.L. Fogli, E. Lisi, A. Marrone and A. Palazzo,
Phys. Lett. B583 (2004) 149; P.C. de Hollanda and A.Y. Smirnov,
Astropart. Phys. 21 (2004) 287.
\item[{20.}] KamLAND: T. Araki et al., arXiv: hep-ex/0406035.
\item[{21.}] A. Bandyopadhyay, S. Choubey, S. Goswami, S.T. Petcov and
D.P. Roy, arXiv:hep-ph/0406328.
\item[{22.}] J.N. Bahcall, M.C. Gonzalez-Garcia and C. Pena-Garay,
JHEP 0406 (2004) 016; O. Miranda, M. Tartola and J.W. Valle, arXiv;
hep-ph/0406280. 
\item[{23.}] A. Bandyopadhyay, S. Choubey and S. Goswami,
Phys. Rev. D67 (2003) 113011.
\end{enumerate}

\newpage

\begin{figure}
\begin{center}
\epsfig{height=8cm,width=8cm,angle=-90,file=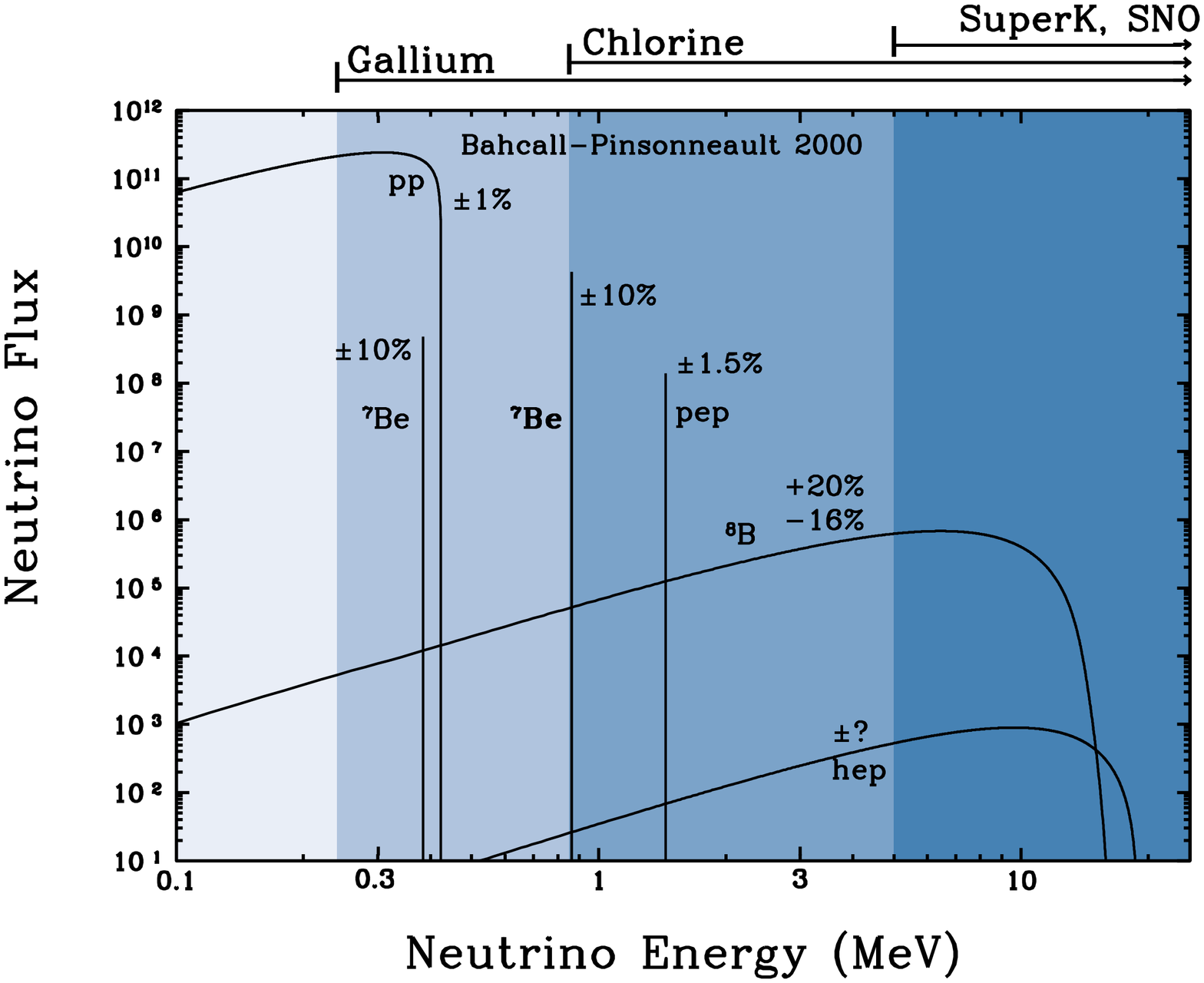}
\caption{The SSM prediction for the solar neutrino fluxes is shown 
along with the energy ranges of the solar neutrino experiments [2].}
\end{center}

\begin{center}
\epsfig{height=8cm,width=8cm,file=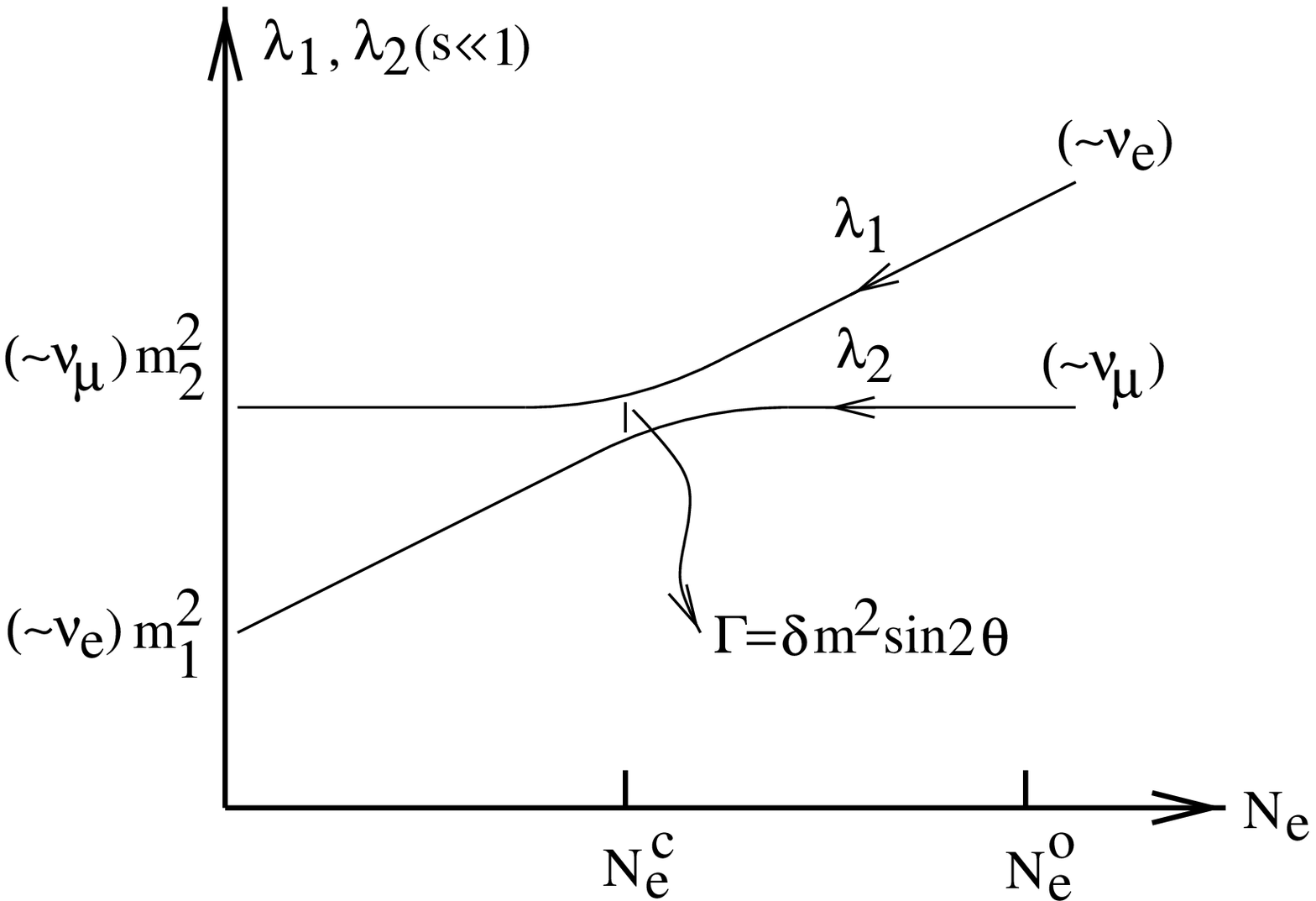}
\caption{Schematic diagram of the effective mass (energy) eigenvalues
of $\nu_{e,\mu}$ as functions of the solar electron density.}
\end{center}
\end{figure}

\newpage

\begin{figure}
\begin{center}
\epsfig{height=7cm,width=7cm,file=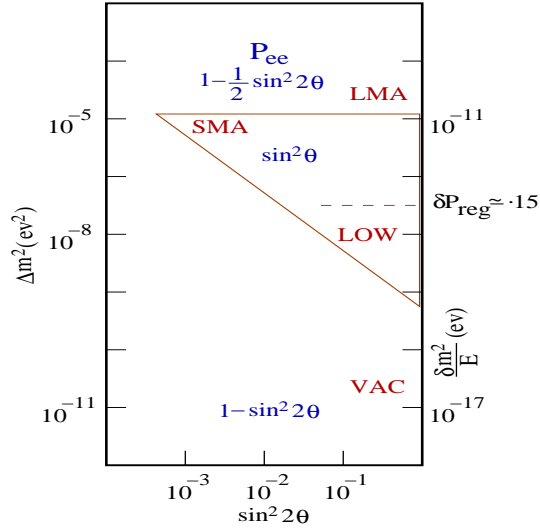}
\caption{The positions of the MSW triangle, the earth regeneration
effect and the vacuum oscillation maximum at $E = 1$ MeV are shown
along with those of the SMA, LMA, LOW and VAC solutions.  While the
former positions scale with $E$ the latter ones are independent of it.}
\end{center}
\medskip
\begin{center}
\epsfig{height=7cm,width=7cm,file=fig4.eps}
\caption{The predicted $\nu_e$ survival probabilities (rates) for the
SMA, LMA, LOW and VAC solutions.}
\end{center}
\end{figure}

\newpage

\begin{figure}
\begin{center}
\epsfig{height=8cm,width=8cm,file=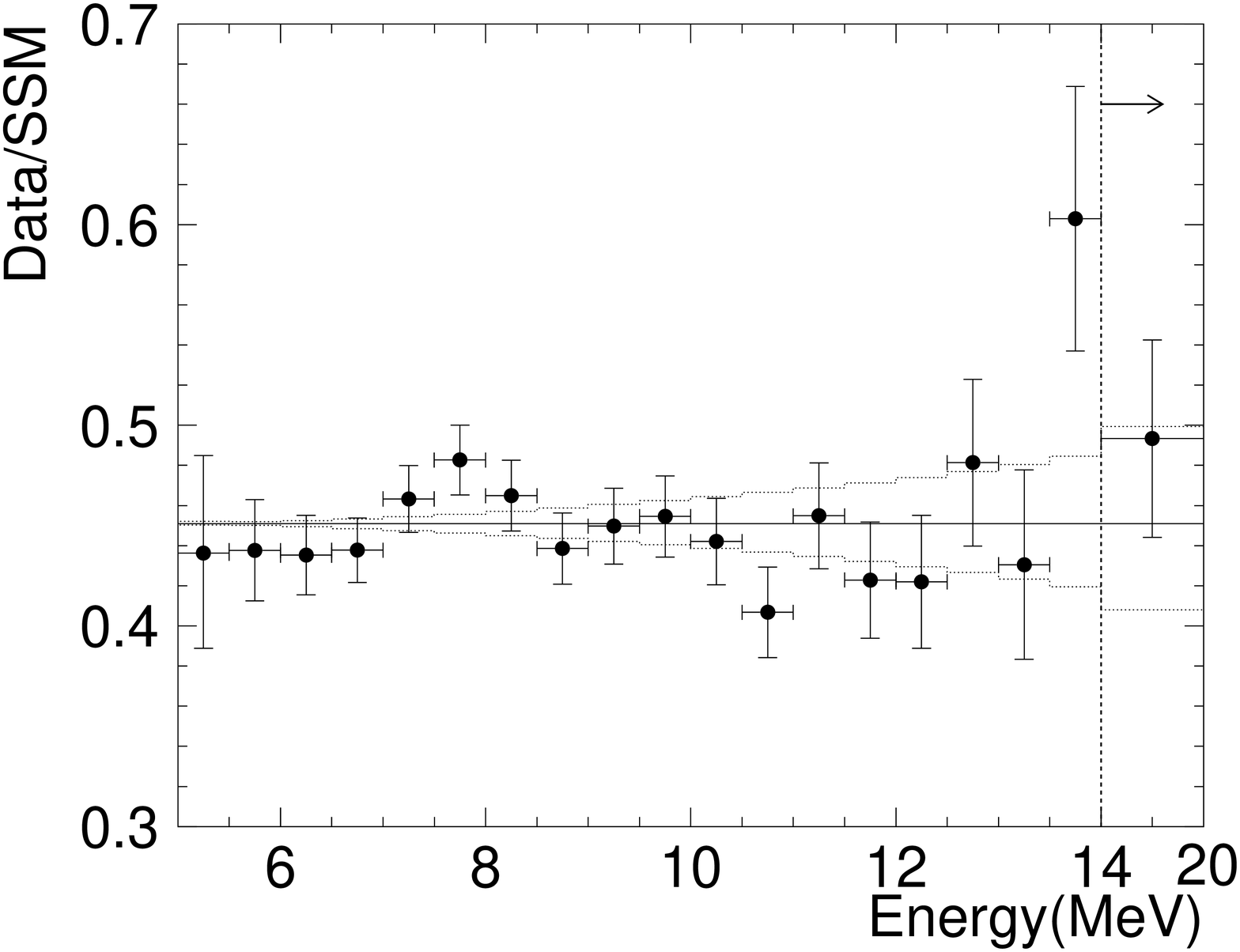}
\caption{The energy (in)dependence of the SK spectrum [5].}
\end{center}

\begin{center}
\epsfig{height=8cm,width=8cm,file=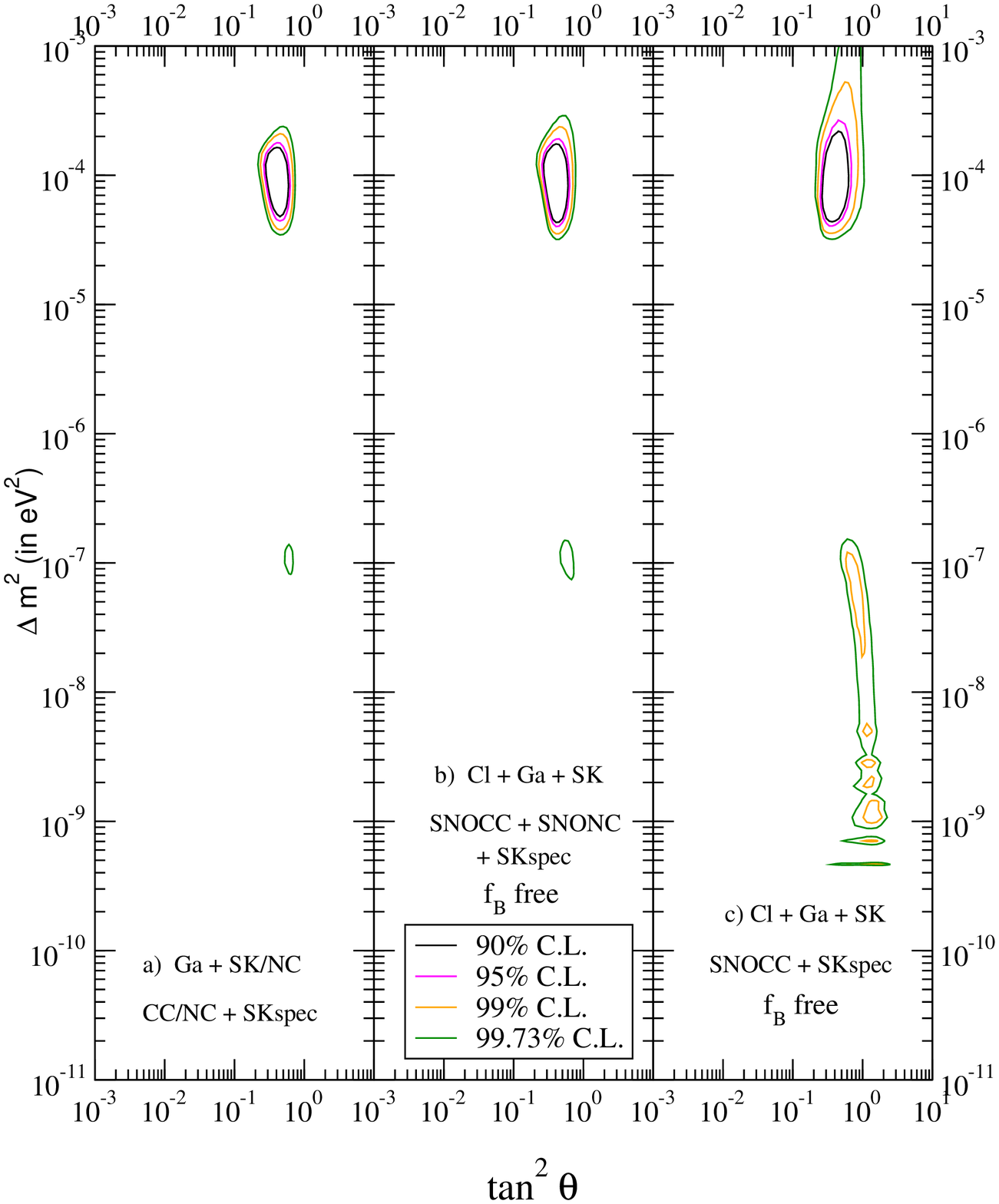}
\caption{Results of global solar neutrino data fits with (left and
middle) and without (right pannel) the first SNO neutral current data [12].}
\end{center}
\end{figure}

\newpage

\begin{figure}
\begin{center}
\epsfig{height=7.5cm,width=7.5cm,file=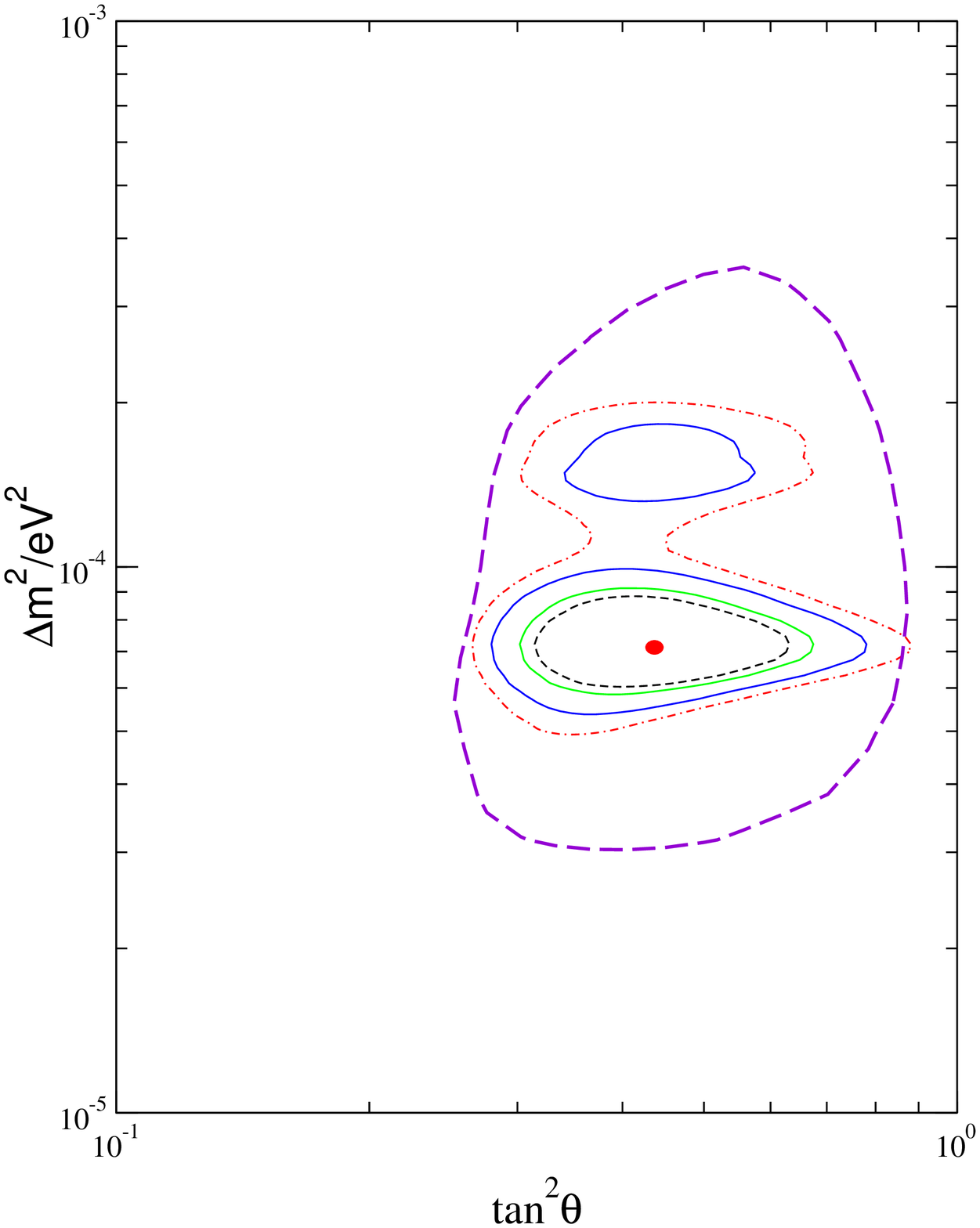}
\caption{The 90,95,99 and 99.73\% CL contours of global fit to solar
neutrino plus the 162 ty KL data.  The 99.73\% CL $(3\sigma)$
contour of the solar fit is shown for comparison [16].}
\end{center}

\begin{center}
\epsfig{height=7.5cm,width=7.5cm,file=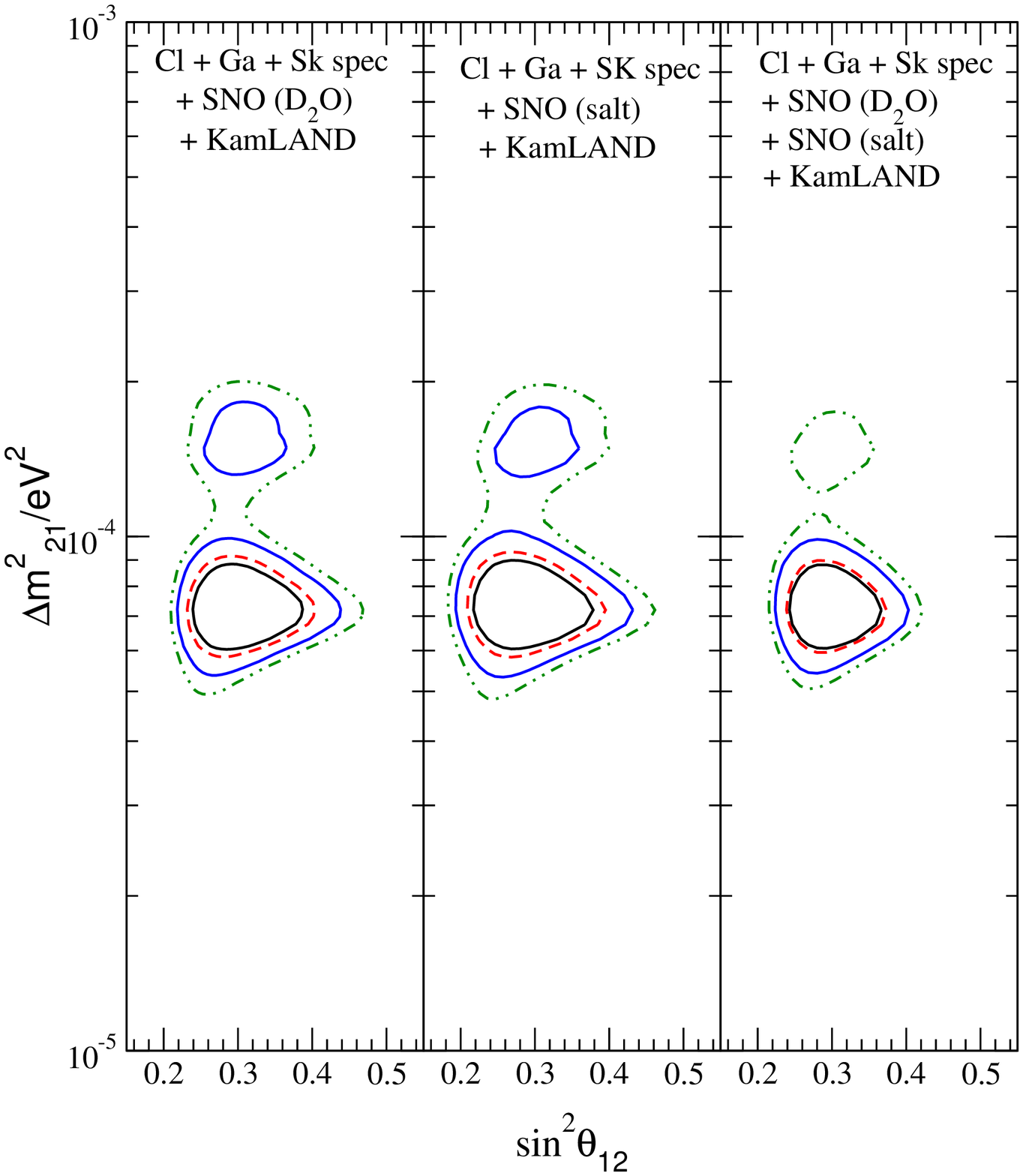}
\caption{The 90,95,99 and 99.73\% CL contours from global fits to
solar neutrino plus 162 ty KL data with phase-I, phase-II and
combined SNO data [18].}
\end{center}
\end{figure}

\newpage

\begin{figure}
\begin{center}
\epsfig{height=7.5cm,width=7.5cm,file=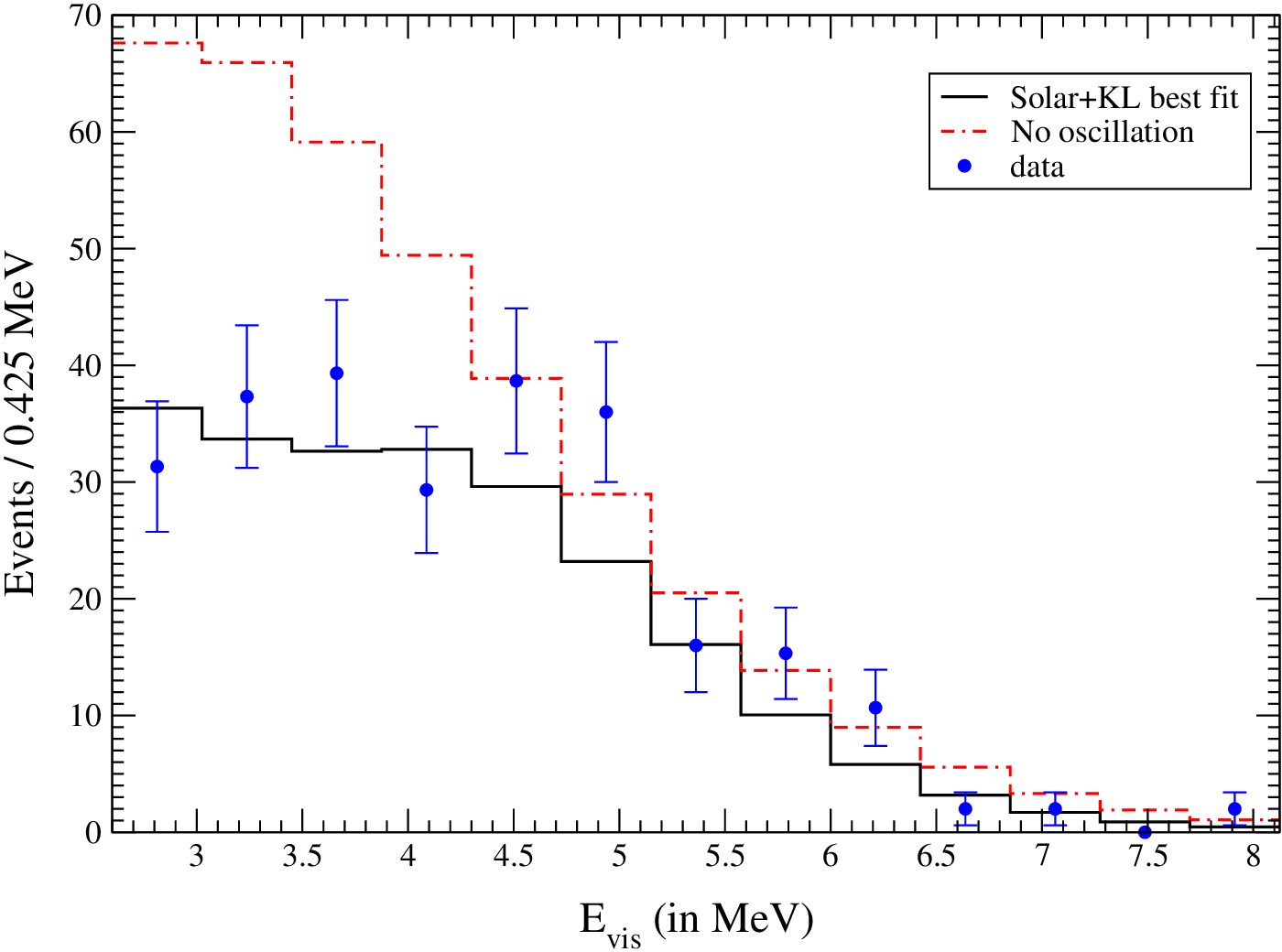}
\caption{The 766 ty KL spectrum compared with the no-oscillation and 
oscillation best fit predictions [21].}
\end{center}
\bigskip
\begin{center}
\epsfig{height=7.5cm,width=7.5cm,file=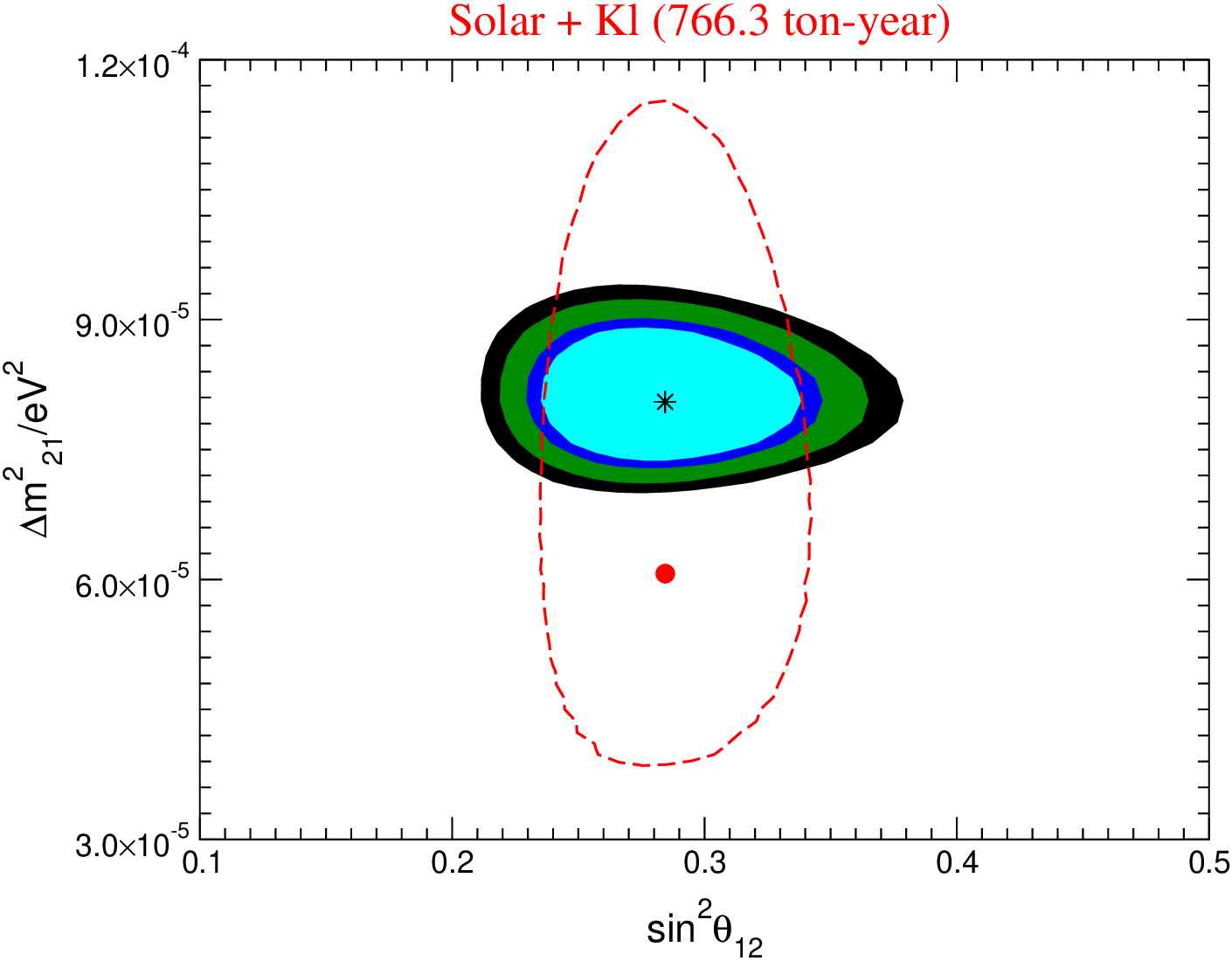}
\caption{The 90,95,99 and 99.73\% contours from global fit to solar
neutrino plus 766 ty KL data. The 90\% CL contour of the solar fit is
shown for comparison [21].}
\end{center}
\end{figure}

\end{document}